\documentclass[aps,prc,twocolumn,floatfix,nofootinbib,showpacs,superscriptaddress]{revtex4-1}
\usepackage{graphicx,amsmath,amssymb,bm}
\usepackage{color}

\newcommand{\tr}{\mathrm{tr}_\mathrm{CD}\!\ }

\begin{document}

\title{Strong $\bm{D^*\to D\pi}$ and $\bm{B^*\to B\pi}$ couplings}

\author{Bruno~El-Bennich}
\affiliation{LFTC, Universidade Cruzeiro do Sul, Rua Galv\~ao Bueno, 868, 01506-000 S\~ao Paulo, SP, Brazil}
\affiliation{Instituto de F\'isica Te\'orica, Universidade Estadual Paulista, Rua Dr.~Bento Teobaldo Ferraz, 271, 01140-070 S\~ao Paulo, SP, Brazil}
\affiliation{Kavli Institute for Theoretical Physics China, CAS, Beijing 100190, China}

\author{Mikhail~A.~Ivanov}
\affiliation{Kavli Institute for Theoretical Physics China, CAS, Beijing 100190, China}
\affiliation{Bogoliubov Laboratory of Theoretical Physics, Joint Institute for Nuclear Research, 141980 Dubna, Russia}

\author{Craig~D.~Roberts\, }
\affiliation{Kavli Institute for Theoretical Physics China, CAS, Beijing 100190, China}
\affiliation{Physics Division, Argonne National Laboratory, Argonne, Illinois 60439, USA [http://www.phy.anl.gov/theory]}
\affiliation{Department of Physics, Peking University, Beijing 100871, China}

\date{22 December 2010}

\begin{abstract}
We compute $g_{D^*\! D\pi}$ and $g_{B^*\! B\pi}$ using a framework in which all elements are constrained by Dyson-Schwinger equation studies of QCD, and therefore incorporates a consistent, direct and simultaneous description of light- and heavy-quarks and the states they may constitute.  We link these couplings with the heavy-light-meson leptonic decay constants, and thereby obtain $g_{D^*\! D\pi}=15.9^{+2.1}_{-1.0}$ and $g_{B^*\! B\pi}=30.0^{+3.2}_{-1.4}$.  From the latter we infer $\hat g_B=0.37^{+0.04}_{-0.02}$.  A comparison between $g_{D^*\! D\pi}$ and $g_{B^*\! B\pi}$ indicates that when the $c$-quark is a system's heaviest constituent, $\Lambda_\mathrm{QCD}/m_c$-corrections are not under good control.
\end{abstract}
\pacs{%
13.25.Ft, 
14.40.Nd,  
11.15.Tk,  
24.85.+p}

\maketitle

\emph{\textbf{Introduction}}.
Non-perturbative QCD effects in heavy flavor physics are a persistent challenge to an accurate determination of the Cabibbo-Kobayashi-Maskawa matrix elements, to which the BaBar, Belle, CDF, CLEO and FOCUS collaborations have dedicated significant effort in the past decade.  Besides improvements in the determination of these Standard Model parameters, the CLEO collaboration also performed a first measurement of the width of the $D$ meson's nearest resonance.  Their reported value for the charged vector meson, $\Gamma (D^{*+} )= 96\pm 4\pm 22\,$keV \cite{Anastassov:2001cw}, is of great
interest because it opens a window on nonperturbative strong physics involving heavy quarks.  More specifically, it allows an extraction of the coupling $g_{D^*\! D\pi}$, which some relate to a putative universal strong coupling, $\hat g$, between heavy-light-vector and -pseudoscalar mesons and a low-momentum pion in a heavy-meson chiral Lagrangian \cite{Casalbuoni:1996pg}.  The step from $g_{D^*\! D\pi}$ to $\hat g$ is contentious, however, because the $c$-quark is not truly heavy and hence corrections to the heavy-quark limit are not necessarily under good control.


In attempting to compute $\hat g$, one may work with the matrix element
\begin{equation}
  \langle  H(p_2) \pi( q) | H^*(p_1,\lambda) \rangle = g_{H^*\!H\pi} \  \bm{\epsilon}_\lambda\!\cdot q \ ,
\label{eq1}
\end{equation}
which defines the dimensionless coupling of a heavy-light vector meson, $H^*$, characterized by a polarization state $\lambda$, and a heavy-light pseudoscalar meson, $H$, to a soft pion with momentum $q=p_1-p_2$.  This matrix element describes the physical processes $D^*\to D\pi$, with both the final pseudoscalar mesons on-shell.  It also serves in computing the unphysical soft-pion emission amplitude $B^*\to B\pi$ in the chiral limit, which defines  $g_{B^* B\pi}$.  A theoretically consistent comparison between these two couplings can provide a quantitative indication of the degree to which notions of heavy-quark symmetry may be applied in the charm sector.

\begin{table}[t!]
\begin{center}
\renewcommand{\arraystretch}{1.4}
\begin{tabular*}
{\hsize}
{
l@{\extracolsep{0ptplus1fil}}
l@{\extracolsep{0ptplus1fil}}
l@{\extracolsep{0ptplus1fil}}
l@{\extracolsep{0ptplus1fil}}
l@{\extracolsep{0ptplus1fil}}}
\hline \hline
 & $g_{D^*\! D\pi}$ & $\hat g_D$ & $g_{B^*\! B\pi}$ & $\hat g_B$  \\  \hline
This work & $15.8_{-1.0}^{+2.1}$ & $0.53_{-0.03}^{+0.07}$  &  $30.0^{+3.2}_{-1.4}$ & $0.37^{+0.04}_{-0.02}$\\\hline
CLEO \protect\cite{Anastassov:2001cw} & $17.9 \pm 1.9$ & $0.61 \pm 0.06$ & & \\ \hline
LQCD$_0^{98}$ \protect\cite{deDivitiis:1998kj}& & & & $0.42\pm 0.09$\\
LQCD$_0^{02}$ \protect\cite{Abada:2002xe} & $18.8\, ^{+2.5}_{-3.0}$ &  $0.67\, ^{+0.09}_{-0.10}$ & & \\
LQCD$_{2}^{08}$ \protect\cite{Ohki:2008py} & & & & $0.52 \pm 0.03$
    \\
LQCD$_{2}^{09}$ \protect\cite{Becirevic:2009xp} & $20\pm 2$ & $0.71 \pm 0.07$ & &  \\
LQCD$_{2}^{09}$ \protect\cite{Becirevic:2009yb} & & & & $0.44 \, ^{+0.08}_{-0.03}$ \\
\hline
SR$^{00}$ \protect\cite{Colangelo:2000dp} & $11 \pm 3$ & $0.36 \pm 0.10 $ & $22\pm 7$ & $0.27 \pm 0.09$\\
SR$^{01}$ \protect\cite{Navarra:2001ju} &  $14 \pm 1.5$  &  $0.47 \pm 0.05 $  &  $42.5 \pm 2.6$ & $0.52 \pm 0.03$  \\
SR$^{06}$ \protect\cite{Duraes:2004uc} &  $17.5 \pm 1.5$  &  $ 0.59 \pm 0.05 $  &  $44.7 \pm 1.0$ & $0.55 \pm 0.01 $  \\

DQM$^{02}$ \protect\cite{Melikhov:2001zv} & $18 \pm 3$ & $0.61 \pm 0.10 $ & $32\pm 5$ & $0.40 \pm 0.06$ \\ \hline \hline
\end{tabular*}
\caption{\label{tablegDsDpi}
Calculated values of $H^\ast \to H \pi$ couplings compared with experiment and other estimates.
For the lattice-QCD results: the subscript indicates the number of dynamical light-fermions employed in the computation;
the valence $c$-quark is treated directly but its dynamics is quenched in all simulations;
and the $B$-meson simulations treat the heavy-quark as static.
NB.\ Where useful, we have combined errors in quadrature in order to simplify presentation. Experimentally \protect\cite{Nakamura:2010zzi} (in GeV):
$m_{D^0}=1.865$,
$m_{D^{\ast +}}=2.010$,
$m_{B^0}=5.280$,
$m_{B^\ast} = 5.325$,
$m_{\pi^+} = 0.1396$, $f_{\pi^+} = 0.1307$.
}
\end{center}\vspace*{-7mm}
\end{table}

Selected results for $g_{H^*\! H\pi}$ and the associated value of
\begin{equation}
\hat g_H := \frac{g_{H^*\! H \pi}}{2 \sqrt{m_H m_{H^\ast}}}\,f_\pi
\end{equation}
are listed in Table~\ref{tablegDsDpi}.  As we explain subsequently, no entirely {\em ab initio} approach to such hadronic decays is currently available.  This might explain the convergence of modern theory for $g_{D^\ast D \pi}$ in the presence of an experimental result but the disagreement over $g_{B^\ast B\pi}$, which is kinematically forbidden.

Present-day simulations of lattice-regularized QCD treat the valence $c$-quark directly as a propagating mode but its dynamics is quenched, whereas the $b$-quark is considered as static.  Beyond statistical, there are errors associated with chiral extrapolation, discretization and perturbative renormalization of the currents involved.  It is apparent from Table~\ref{tablegDsDpi} that results obtained within this approach exhibit a significant difference between $\hat g_D$ and $\hat g_B$.  This is probably because $c$-quark physics is not well-approximated by the heavy-quark limit and hence related observables contain significant $\Lambda_\mathrm{QCD}/m_c$ corrections.

QCD sum-rules, too, have been used to estimate $g_{H^* H\pi}$.  In advance of the CLEO measurement, there were numerous results, which differed amongst themselves by as much as a factor of two; and all of which were more than 25\% below the experimental value determined subsequently \cite{Colangelo:2000dp}.  The value obtained in this approach depends heavily on the functional forms used for extrapolation to the mesons' on-shell momenta from the domains upon which the calculations are actually valid \cite{Navarra:2001ju}.  Based on Ref.\,\cite{Navarra:2001ju}, loop corrections in $D$-meson effective field theory have been used to constrain extrapolation of the $D^*D\pi$ coupling to physical momenta \cite{Duraes:2004uc}.

The simultaneous computation of both couplings has been completed in a dispersive quark model \cite{Melikhov:2001zv}.  However, this approach makes no qualitative distinction between light- and heavy-quarks.  Hence it cannot veraciously describe dynamical chiral symmetry breaking (DCSB), which makes problematic its treatment of the structure and interactions of pions.

Symmetry-preserving models built upon robust predictions of QCD's Dyson-Schwinger equations (DSEs) \cite{Maris:2003vk,Roberts:2007jh,Roberts:2007ji} provide a sound framework within which to examine heavy-meson observables \cite{Ivanov:1997yg,Ivanov:1997iu,Ivanov:1998ms,Ivanov:2007cw}.  Such studies describe quark propagation by fully dressed Schwinger functions.  This has a material impact on light-quark characteristics \cite{Roberts:2007ji}.  Constrained by experimental and theoretical heavy-light-meson information available at the time, Ref.\,\cite{Ivanov:1998ms} obtained $g_{D^*\! D\pi}=11$, $\hat g_D=0.37$ and $g_{B^*\! B\pi}=23$, $\hat g_B=0.29$.  These values compare well with a representative average of theoretical estimates then available; viz., $g_{D^*\! D\pi} =12\pm 4$, $g_{B^*\! B\pi}=25\pm 7$ \cite{Belyaev:1994zk}.  However, $g_{D^*\! D\pi}=11$ is not consistent with the subsequent CLEO measurement \cite{Anastassov:2001cw}.  Numerous advances have been made in experiment and theory in the intervening period.  Herein, therefore, we reassess the DSE study \cite{Ivanov:1998ms}, implement improvements detailed in Ref.\,\cite{Ivanov:2007cw} and add one more; namely, an expression of the probable difference in size between the vector and pseudoscalar heavy-light mesons.

\begin{figure}[t]
\begin{center}
\includegraphics[width=0.28\textwidth]{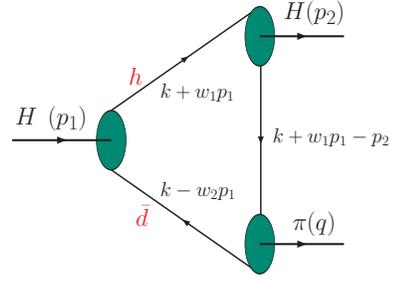}
\caption{\label{Fig1}
(Color online) Pictorial representation of Eq.\,\eqref{eq2}, our impulse approximation to the \mbox{$H^\ast \to H \pi$} decay: \emph{solid lines} -- dressed-quark
propagators [Eqs.\,(\ref{ssm}), (\ref{svm}), (\ref{SQ})]; and \emph{filled ellipses} -- meson Bethe-Salpeter amplitudes [Eqs.\,(\ref{piKamp}) -- (\protect\ref{GHs})].}
\end{center}
\end{figure}

%
\emph{\textbf{Framework and tools}}.
At leading-order in a systematic, nonperturbative, symmetry-preserving DSE truncation scheme \cite{Munczek:1994zz,Bender:1996bb}, the $H^\ast \to H \pi$ decay amplitude is given by
\begin{eqnarray}
  g_{H^*\!H\pi} \  \bm{\epsilon}^\lambda\!\cdot q  & =  &  \tr \! \int\! \frac{d^4k}{(2\pi)^4} \, \bm{\epsilon}^\lambda\cdot \Gamma_{H^*}(k;p_1)  S_Q(k_Q) \nonumber   \\
  &  &\hspace*{-1cm} \times \  \bar  \Gamma_H(k;-p_2) S_f(k_f^\prime) \bar \Gamma_\pi(k;-q) S_f(k_f) \ ,
\label{eq2}
\end{eqnarray}
where: the trace is over color and spinor indices;
$k_Q = k+ w_1 p_1$, $k_f^\prime = k + w_1 p_1 - p_2$, $k_f= k- w_2 p_1$, and the relative-momentum partitioning parameters satisfy $w_1+w_2=1$;
$\bm{\epsilon}^\lambda_\mu$ is the vector-meson polarization four-vector; and $S$ and $\Gamma$, described below, are dressed-quark propagators and meson Bethe-Salpeter amplitudes, respectively.  The approximation represented by the diagram in Fig.\,\ref{Fig1} has been widely used successfully; e.g., Refs.\,\cite{Roberts:1994hh,Jarecke:2002xd,Maris:2003vk,Holl:2005vu,Roberts:2007jh}.  It is reasonable to expect that corrections associated with final-state interactions are small owing to the large $c$-quark mass.

Along with the radiative-pion couplings, we simultaneously calculate the $H^\ast$- and $H$-meson leptonic decay constants, which are determined via \cite{Ivanov:1998ms,Maris:1997hd,Maris:1997tm}:
\begin{eqnarray}
\label{psdecay}
P_\mu f_{H}  &=& \tr  \int\!  \frac{d^4k}{(2\pi)^4} \, \gamma_5 \gamma_\mu\, \chi_{H}(k;P)\,, \\
\label{vecdecay}
M_{H^\ast} f_{H^\ast} & = & \frac{1}{3}\tr  \int\!  \frac{d^4k}{(2\pi)^4} \,  \gamma_\mu\, \chi_{\mu H^\ast}(k;P) \, ,
\end{eqnarray}
where $\chi(k;P) = S_{f_1}(k+w_1P) \Gamma(k;P) S_{f_2}(k-w_2P)$.  The Bethe-Salpeter amplitudes are canonically
normalized; i.e.,
\begin{eqnarray}
  2\, P_\mu & = & \left [ \frac{\partial}{\partial K_\mu} \Pi(P,K) \right ]_{K=P}^{P^2=-m^2_{0^-}} \ ,
  \label{norm1} \\
   \Pi(P,K) & = &   \tr \int\!  \frac{d^4k}{(2\pi)^4} \, \bar \Gamma_{0^-}(k;-P) S_{f_1}(k+w_1K) \nonumber \\
   &  & \times \   \Gamma_{0^-}(k;P) S_{f_2}(k-w_2K) \,,
  \label{norm2}
\end{eqnarray}
with an analogous expression for the $H^\ast$.

For any quark flavor, the dressed-quark propagator has the general form
\begin{equation}
S(p) = -i \gamma\cdot p\, \sigma_V(p^2) + \sigma_S(p^2)
= 1/[i\gamma\cdot p\, A(p^2) + B(p^2)], \label{SpAB}
\end{equation}
and can be obtained from QCD's gap equation \cite{Maris:2003vk}.  In connection with light-quarks, it is a longstanding prediction of DSE studies that both the wave function renormalization, $Z(p^2)=1/A(p^2)$, and dressed-quark mass, $M(p^2)=B(p^2)/A(p^2)$, receive strong momentum-dependent corrections at infrared momenta: $Z(p^2)$ is suppressed and $M(p^2)$ enhanced.  These features are an expression of DCSB and, plausibly, of confinement.\footnote{Equations\,(\protect\ref{ssm}), (\protect\ref{svm}) represent $S(p)$ as an entire function.  This entails the absence of a Lehmann representation, which is a sufficient condition for confinement \protect\cite{Roberts:2007ji}.} The enhancement of $M(p^2)$ is central to the appearance of a constituent-quark mass-scale and an existential prerequisite for Goldstone modes.   These DSE predictions are confirmed in numerical simulations of lattice-QCD \cite{Zhang:2004gv}.

The impact of this infrared dressing on hadron phenomena has long been emphasized \cite{Roberts:1994hh} and, while numerical solutions of the quark DSE are now readily obtained, the utility of an algebraic form for $S(p)$ when calculations require the evaluation of multi-dimensional integrals is self-evident.  An efficacious parametrization, which exhibits the features described above, has been used extensively; e.g., \cite{Ivanov:1998ms,Ivanov:2007cw,Hecht:2000xa,Cloet:2008re}.  It is expressed via
\begin{eqnarray}
\nonumber \bar\sigma_S(x) & =&  2\,\bar m \,{\cal F}(2 (x+\bar m^2))\\
&&  + {\cal
F}(b_1 x) \,{\cal F}(b_3 x) \,
\left[b_0 + b_2 {\cal F}(\epsilon x)\right]\,,\label{ssm} \\
\label{svm} \bar\sigma_V(x) & = & \frac{1}{x+\bar m^2}\, \left[ 1 - {\cal F}(2 (x+\bar m^2))\right]\,,
\end{eqnarray}
with $x=p^2/\lambda^2$, $\bar m$ = $m/\lambda$, ${\cal F}(x)= [1-\exp(-x)]/x$,
$\bar\sigma_S(x) = \lambda\,\sigma_S(p^2)$ and $\bar\sigma_V(x) =
\lambda^2\,\sigma_V(p^2)$.  The parameter values were fixed in Ref.\,\cite{Ivanov:1998ms} by requiring a least-squares fit to a wide range of light- and heavy-meson observables, and take the values:\footnote{$\epsilon=10^{-4}$ in Eq.\,(\protect\ref{ssm}) acts only to
decouple the large- and intermediate-$p^2$ domains \protect\cite{Roberts:1994hh}.}
\begin{equation}
\label{tableA}
\begin{array}{llcccc}
f &   \bar m_f& b_0^f & b_1^f & b_2^f & b_3^f \\\hline
u=d &   0.00948 & 0.131 & 2.94 & 0.733 & 0.185
\end{array}\;.
\end{equation}
The mass-scale $\lambda=0.566\,$GeV, with which value the current-quark mass is $m_d=5.4\,$MeV.  In addition one obtains the following Euclidean constituent-quark mass, defined as
$\hat M^2 = \{s|s+M^2(s)=0\}$: $\hat M_d = 0.36\,$GeV.

As noted elsewhere \cite{ElBennich:2008xy,ElBennich:2008qa}, studies which do not or cannot implement light-quark dressing in this QCD-consistent manner, invariably encounter problems because of the need to employ rather large constituent-quark masses and the associated poles in the light-quark propagators.  This translates into considerable model-sensitivity in the results for any heavy-light form factors, such
as $B$- to light-meson transition form factors \cite{ElBennich:2009vx}.

Whilst the impact of DCSB on light-quark propagators is significant, it is less so for heavier quarks.  This is plain in Fig.~1 of Ref.\,\cite{Ivanov:1998ms}.  It can also be seen by considering the renormalization-point-invariant ratio $\varsigma_f:=\sigma_f/M^E_f$, where $\sigma_f$ is a constituent-quark $\sigma$-term \cite{Roberts:2007jh}.  This ratio measures the effect of explicit chiral symmetry breaking on the dressed- quark mass-function compared with the sum of the effects of explicit and dynamical chiral symmetry breaking.  Calculation reveals \cite{Roberts:2007jh}
$\varsigma_d = 0.02$, $\varsigma_s = 0.23$, $\varsigma_c = 0.65$, $\varsigma_b = 0.8$,
results which are readily understood.  Naturally, $\varsigma_f$ vanishes in the chiral limit and must be small for light-quarks because the magnitude of their constituent-mass owes primarily to DCSB.  For heavy-quarks, $\varsigma_f$ approaches unity because explicit chiral symmetry breaking becomes the dominant source of their mass.  We therefore use a constituent-quark-like propagator for $c$- and $b$-quarks; viz.,
\begin{equation}
\label{SQ}
S_{Q}(k) = \frac{1}{i \gamma\cdot k + \hat M_Q}\,,\; Q=c,b,
\end{equation}
where \cite{Ivanov:1998ms}:
$\hat M_c = 1.32\,${\rm GeV}, 
$\hat M_b = 4.65\,${\rm GeV}.

%
The meson Bethe-Salpeter amplitudes, which appear in Fig.\,\ref{Fig1} and are consistent with the generalized impulse approximation, are properly obtained from an improved-ladder Bethe-Salpeter equation \cite{Maris:2003vk}.  The solution of this equation requires a simultaneous solution of the quark DSE.  However, since we have already chosen to simplify the calculations by parametrizing $S(p)$, we follow Ref.~\cite{Ivanov:1998ms} and also employ that expedient with $\Gamma$.
The axial-vector Ward-Takahashi identity and DCSB have an enormous impact on the structure and properties of light pseudoscalar mesons.  Indeed, the quark-level Goldberger-Treiman relations derived in Ref.\,\cite{Maris:1997hd} motivate and support the following efficacious parametrization of the pion's Bethe-Salpeter amplitude:
\begin{equation}
\label{piKamp}
\Gamma_{\pi}(k;P) = i\gamma_5\,\frac{\surd 2}{f_{\pi}}\,B_{\pi}(k^2)\,,
\end{equation}
where $B_\pi:=\left.B_u\right|_{b_0^u\to b_0^\pi}$ is obtained from Eqs.\,(\ref{SpAB}) -- (\ref{svm}) through the replacement $b_0^u \rightarrow b_0^{\pi}=0.204$ \cite{Ivanov:1998ms}.  Equation~(\ref{piKamp}) expresses an intimate connection between the leading covariant in a pseudoscalar meson's Bethe-Salpeter amplitude and the scalar piece of the dressed-quark self-energy \cite{Maris:1997hd}.

Whilst the renormalization-group-improved rainbow-ladder DSE kernel is appropriate for the study of mesons constituted from equal-mass constituents \cite{Maris:2003vk,Bhagwat:2006xi}, this is not so for heavy-light mesons.  In such systems cancelations, which largely mask the effect of dressing the quark-gluon vertices, are blocked by the dressed-propagator asymmetry; e.g., a recent analysis \cite{Nguyen:2009if} obtained a fair description of $D$- and $B$-meson masses but underestimated their leptonic decay constants by 30-50\%.  (The approach introduced in Ref.~\cite{Chang:2009zb} might be the advance needed to make progress with heavy-light systems.)
Moreover, as we have already chosen to simplify the calculations by parametrizing $S(p)$ and $\Gamma_\pi$, it is rational to follow Refs.\,\cite{Ivanov:1998ms,Ivanov:2007cw} and employ that expedient, too, with $\Gamma_{H^\ast,H}$:
\begin{eqnarray}
\label{GH}
    \Gamma_H (k;P) &=& i \gamma_5 \, \frac{\exp (-k^2/\omega_H^2) }{\mathcal{N}_H}  \,, \\
\label{GHs}
   \bm{\epsilon}^\lambda \cdot \Gamma_{H^*} (k;P) &=&  \bm{\epsilon}^\lambda\cdot \gamma  \, \frac{\exp (-k^2/ \omega_{H^*}^2) }{\mathcal{N}_{H^*}}  \,.
\end{eqnarray}
The normalization, $\mathcal{N}_{H^{(*)}}$, is fixed by Eqs.~\eqref{norm1} and \eqref{norm2}.  Herein, however, we depart from Refs.~\cite{Ivanov:1998ms,Ivanov:2007cw} and do not assume heavy-quark symmetry to be realized exactly; namely, we eschew the spin-independent \emph{Ansatz} and fit each of the parameters $\omega_H$ and $\omega_{H^*}$  to their respective leptonic decay constants, either known from experiment or estimated via lattice-simulations.  [NB.\ Our results are not materially affected by the pointwise form of the functions in Eqs.\,(\ref{GH}), (\ref{GHs}).]

Poincar\'e invariance is a feature of the direct application of DSEs to the calculation of hadron properties.  However, that is compromised if one does not retain the complete structure of hadron bound-state amplitudes \cite{Maris:1997tm}.  This applies herein because we use one-covariant models for the amplitudes.
%
%
To proceed we must therefore specify the relative momenta.

As indicated in Fig.~\ref{Fig1}, in computing $g_{H^\ast H \pi}$: we allocate a fraction $w_1$ of the heavy-light-meson's momentum to the heavy-quark; $w_2$ to the light-quark; and momentum-conservation specifies the remaining momentum.
An analogous procedure is followed for the leptonic decays and normalization, Eqs.\,\eqref{psdecay} -- \eqref{norm2}.
The relative-momentum partitioning parameter is defined via a center-of-mass prescription; viz.,
\begin{equation}
w_1^h = \frac{\hat M_h}{\hat M_h + \hat M_d}\,,
\end{equation}
which yields  $w_1^c =0.79$, $w_1^b = 0.93$.
In addition, we interpret the amplitudes $\Gamma$ in Eqs.\,(\ref{piKamp}) -- (\ref{GHs}) as the zeroth Chebyshev moment of the leading-covariant in the relevant meson's Bethe-Salpeter amplitude.  We emphasize that any sensitivity to a definition of the relative momenta is an artifact owing to our simplifications.  \emph{Every} study that fails to retain the full structure of the Bethe-Salpeter amplitude shares this complication, which is never encountered in complete studies; e.g.,  Refs.~\cite{Maris:1997tm,Bhagwat:2006xi,Nguyen:2009if}.

\emph{\textbf{Results}}.
%
In Ref.\,\cite{Ivanov:1998ms} it was noted that it is a poor approximation to assume heavy-quark symmetry for $c$-quark systems because corrections can be as large as a factor of two in semileptonic $c\to d$ transitions.  This is also seen in Ref.~\cite{Ivanov:2007cw}, wherein the assumption $\omega_{D^*}=\omega_D$ yields $f_{D^\ast}-f_D=321-223=98\,$MeV, whereas lattice-QCD combined with experiment suggests a smaller difference: $39\pm 29\,$MeV.  In relation to $b$-quark mesons, whilst $\omega_{B^*}=\omega_B$ is more likely to be a good approximation, it cannot be exact and hence it is worth exploring the impact of relaxing this constraint.

\begin{figure}[t!]
\begin{center}
\includegraphics[width=0.43\textwidth]{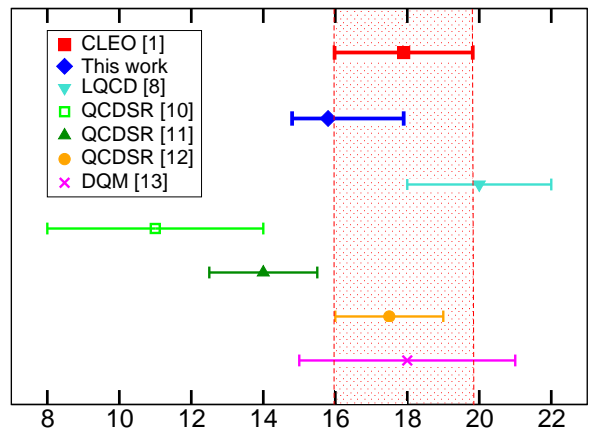}
\caption{\label{Fig2}
(Color online) Dimensionless coupling $g_{D^*\! D\pi}$: comparison between experiment and recent estimates, with their associated errors added in quadrature.  (See Legend and Table~\ref{tablegDsDpi} for details.)}
\end{center}
\end{figure}

We therefore have four parameters: $\omega_D$, $\omega_{D^*}$, $\omega_B$ and $\omega_{B^*}$, which we determine via a least-squares fit to the leptonic decay constants [in MeV]:
$f_D= 205.8 \pm 8.9$ \cite{Eisenstein:2008sq};
$f_{D^\ast} = 245\pm 20$ \cite{Becirevic:1998ua};
$f_{B} = 229 ^{+51}_{-48}$ \cite{Ikado:2006un};
$f_{B^\ast} = 196 ^{+46}_{-24}$ \cite{Becirevic:1998ua},
using Eqs.\,\eqref{psdecay} and \eqref{vecdecay}.  A perfect fit is possible, and is obtained with [in GeV]:
\begin{equation}
\begin{array}{rclrcl}
  \omega_D &=& 1.26 \pm 0.10 \,, \;    &     \omega_{D^*} & = & 1.21 \pm 0.14 \, , \\
  \omega_B &= & 2.26 \; ^{+0.76}_{-0.68} \, , \;   & \omega_{B^*} & = &  1.58 \; ^{+0.52}_{-0.27}\,.
\label{decayconstants}
\end{array}
\end{equation}
It is notable that, in contrast to Ref.\,\cite{Ivanov:2007cw}, the width parameters are all consistent with intuition:
\mbox{$\ell_{\omega_B}:=1/\omega_B = 0.09\,$fm}
$\,<\ell_{\omega_B^\ast}=0.12\,$fm
$\,<\ell_{\omega_D}=0.157\,$fm
$\,<\ell_{\omega_D^\ast}=0.163\,$fm; namely, by this rudimentary gauge, pseudoscalar mesons are smaller than vector mesons and systems containing a single $b$-quark are smaller than those containing a $c$-quark.

Using the width parameters in Eq.~(\ref{decayconstants}), we computed the strong $H^{*+} \! H^0\pi^+$ couplings
\begin{equation}
g_{D^*\! D\pi} = \lim_{q^2\to -m_\pi^2}\,  g_{D^*\! D\pi} (q^2) \,,\;
g_{B^*\! B\pi} = \lim_{q^2\to 0} \, g_{B^*\! B\pi} (q^2) \,,
\end{equation}
where the latter serves as a definition since the process $B^\ast \to B \pi$ is kinematically forbidden.  We list our results in Table~\ref{tablegDsDpi} and, for $g_{D^*\! D\pi}$, present a pictorial comparison with experiment and other studies in Fig.\,\ref{Fig2}.  Importantly, in our approach one can directly calculate the amplitude at the fully on-shell point and with the physical light-quark current-mass: \emph{no} extrapolations are necessary.

There are two obvious sources of uncertainty in our results for $g_{H^*\! H\pi}$.
The errors in the values of the leptonic decay constants translate into the uncertainties quoted on the width parameters in Eq.\,\eqref{decayconstants} and produce a range of values for $g_{H^*\! H\pi}$.
In addition, a change of $\pm 20$\% in $w_2^{c,b}$ gives variations in $g_{D^*\! D\pi}$ and $g_{B^*\! B\pi}$ of $\sim 10$\% and $\sim 5$\%, respectively.
We treat these variations as independent uncertainties on each coupling and add them in quadrature to produce the errors quoted in Table~\ref{tablegDsDpi}.

Our results are significantly different to those reported in Ref.\,\cite{Ivanov:1998ms} because, as noted above, there are two material differences between this calculation and that.  Namely, we allow: $\omega_{D^\ast} \neq \omega_D$ and $\omega_{B^\ast} \neq \omega_B$; and the light-quark to carry a fraction of the $H^\ast$-meson's momentum.  In connection with $g_{B^*\! B\pi}$, since the $b$-quark is genuinely heavy and should therefore carry most of the $B^\ast$-meson's momentum, the improvement arises primarily because $\omega_{B^\ast} \neq \omega_B$.  On the other hand, the $c$-quark is neither light nor truly heavy and thus $g_{D^*\! D\pi}$ is quite sensitive to the amount of the $D^\ast$-meson's momentum carried by the light-quark.  If we require $w_2^c=0$, as in Ref.\,\cite{Ivanov:1998ms}, then one finds $g_{D^*\! D\pi}<13$, even allowing for $\omega_{D^\ast} \neq \omega_D$.
A less important factor is our use of updated $D$, $D^*$, $B$ and $B^*$ masses.  These values influence the normalization constants, Eq.\,\eqref{norm2}, but not to an extent which requires the effect of experimental mass uncertainties to be included in our error estimate for $g_{H^*\! H\pi}$.

\emph{\textbf{Summary and Conclusions}}.
We presented a calculation of $g_{D^*\! D\pi}$ and $g_{B^*\! B\pi}$ based upon QCD's Dyson-Schwinger equations (DSEs).  By implementing a more realistic representation of heavy-light mesons; e.g., allowing the light-quark in the heavy-light-meson to carry some of the meson's momentum and for a difference between the sizes of pseudoscalar and vector mesons, our analysis improves significantly upon earlier DSE-based studies.
Furthermore, we step beyond other models because, by expressing confinement and dynamical chiral symmetry breaking (DCSB) in a manner compatible with the predictions of QCD's DSEs, our approach incorporates a consistent, direct and simultaneous description of light- and heavy-quarks and the states they may constitute.
Finally, even with respect to modern numerical simulations of lattice-QCD, our approach has merits; e.g., direct access to the chiral limit, a veracious expression of DCSB and the reliable treatment of light-quarks, and a dynamical treatment of all quarks.

Quantitatively, our study links the leptonic decay constants of heavy-light-mesons and their radiative-pion decays.  This connection provides a natural explanation of the experimental value for $g_{D^*\! D\pi}$ and a prediction for the putative universal strong coupling, $\hat g$, between heavy-light-vector and -pseudoscalar mesons.  In this connection our results emphasize that when the $c$-quark is a system's heaviest constituent, $\Lambda_\mathrm{QCD}/m_c$-corrections are not under good control.
One should be mindful of this when estimating, e.g., the kinematically forbidden couplings between $D$- and light-vector-mesons that are used in phenomenological models of charmonium production.

\emph{Acknowledgments}.
B.~El-Bennich acknowledges the hospitality of staff at the Bogoliubov Laboratory of Theoretical Physics where this work was initiated.
Work supported by:
Funda\c{c}\~ao de Amparo \`a Pesquisa do Estado de S\~ao Paulo, grant nos.~2009/51296-1 and 2010/05772-3;
Russian Fund for Basic Research grant No.~10-02-00368-a;
and
the United States Department of Energy, Office of Nuclear Physics, contract no.~DE-AC02-06CH11357


\end{document}